\documentclass{article}
\AtBeginDocument{%
  }

\begin{document}

\title{The Divine Software Engineering Comedy\\Inferno: The Okinawa Files}

\author{Michele Lanza}



%

%

\maketitle

\begin{abstract}
In June 2024 I co-organized the FUSE (FUture of Software Engineering) symposium in Okinawa, Japan. Me, Andrian Marcus, Takashi Kobayashi and Shinpei Hayashi were general chairs, Nicole Novielli, Kevin Moran, Yutaro Kashiwa and Masanari Kondo were program chairs, some members of my group (Carmen Armenti, Stefano Campanella, Roberto Minelli) were the tables, can’t have a room with only chairs, after all.
  
We invited a crowd of people to discuss what future software engineering has. FUSE became a 3-day marathon on whether there is actually a future at all for SE.

This essay is a slightly dark take about what I saw at that event, very loosely based on the discussions that took place, adding some healthy sarcasm and cynicism, the intellectual salt and pepper I never seem to run out of. I listened to the brilliant people who gathered to talk about where we're headed, and distilled three nightmares headed in our direction: software makers who don't know what they're doing (but get the job done anyway), a field moving so fast it can't remember its own lessons, and technologies multiplying like rabbits in spring.

So, let's start. The future, eh? The future of software engineering looks like a car crash in slow motion: you can see it coming but you can't look away. The thing is...
\end{abstract}

\section{Prologue - Into the Lungs of Hell}

...the thing about the future is this: it arrives whether you're ready or not.

FUSE 2024 brought together researchers from around the world to talk about software engineering. We met in Okinawa, Japan, at OIST (the Okinawa Institute of Science and Technology), in rooms that smelled of coffee and dry-erase markers, and we talked. Boy, did we talk. Some of us had been in the field for thirty years. Some were young enough to have learned programming from YouTube, pardon: TikTok, videos.

AI dominated every conversation, the way cancer dominates silent dinner conversation when someone's sick. But there are other problems, too. Problems that have been festering for years, growing in the dark like mushrooms in a basement.

Our field was always moving fast, now it is moving too fast. Technologies are multiplying. Education can’t keep up. And somewhere in all of this, the meaning of "software engineering" itself is changing into something difficult to grasp, something eerily slippery.

The Future of Software Engineering? Abandon all hope ye who enter here.

\section{The New Wave of Software Makers}

Here's a story that'll give you the shivers: there's a new kind of person building software now. They're called software makers, and they are a scary mob.

These makers never went to college for computer science. They seem to live in YouTube shorts and TikTok. They never learned about algorithms or data structures. Never stayed up until three in the morning debugging memory leaks. Never studied the dos and do nots, never learned about design patterns, and refactoring, and good style. They vibe code. They code surf. They use AI tools and low-code platforms with funky names. They click some buttons, drag some boxes around, and boom: software that actually works. Kinda. Good enough, anyway.

They work nine to five, no overtime, no weekend hackathons. When something breaks (and things always break, don't they?) they're lost. That's when the real engineers have to show up, swooping in like a SWAT team for code crises, explaining what went wrong and how to fix it.

This is what might be called "fast fashion software." Built quick, built cheap, built to solve today's problem without a thought for tomorrow. Each new task gets brand new code. The old code sits there, gathering dust and technical debt. Quality goes down, down, down. But it works. It solves the problem. The metrics look good. The bosses and the venture capitalists are happy.

Some people think code itself is going to disappear. Not just code as we know it, all of it! Instead, we'll have intent-based programming. You tell the computer what you want in plain English, and some foundation model figures out how to make it happen. The model becomes the processor. You give it intents, it gives you behavior.

This makes requirements engineering important again, after decades of being the field nobody wanted to talk about at the ICSE social events. The problem is always the same: figuring out what people actually want. Humans are terrible at this. We need a new language, something between English and code. Something structured enough that machines can parse it. Something clear enough that humans can read it without a decoder ring.

\section{The Evolution Crisis}

Now we come to the dark stuff. Software evolution research, the study of how software changes over time, is having an identity crisis. Does the field even have a future? Did it ever have a past? Back in the seventies, this guy named Lehman came up with eight laws of software evolution. Eight laws from studying one system. One. In the decades since, we've had thousands of systems, millions of developers, billions of lines of code. How many new laws have we discovered? Zero.

The problem runs deeper than you think. The field moves too fast to learn anything. We, the so-called software engineering research community, publish papers like a factory pumps out widgets. Mine data. Generate statistics. Claim victories. But we're not learning. We're just... producing. CSRankings beckons, after all. No time for any substance, move on, we must appear.

Mining papers are the worst. They crunch numbers but strip away all the context, ignoring the why behind the what. Reviewers want generalization, not case studies. They want universal laws from a field that can't even understand its specific instances. It's like trying to write a unified theory of physics without ever doing a single experiment.

Here's the thing: software is a system of systems. It's people and machines in twisting feedback loops. Evolution means understanding how these systems respond to change. How they adapt. How they fail. But we don't do that work. Industry people know things (real, useful things!), but academia never validates their knowledge. Academics write papers about software they've never actually built.

They're like food critics who've never cooked a meal.

There's another problem, too. We slice our analysis too thin. Look at a project for three months (tops) and call it research. We need bigger samples. Longer timescales. We need negative results, papers about what didn't work and why. But nobody wants to publish those. Failure is something you bury in a drawer and try to forget. The field needs to slow down. To watch. To learn. To remember. But it won't. Can't. The grant cycles won't allow it. The tenure committees won't allow it. The whole machine keeps churning, faster and faster, learning nothing.

Legacy systems will keep evolving, shambling forward like zombies in a bad movie. Too complex to kill. Too essential to ignore. But even fast fashion software teaches lessons if you're willing to look. Why did we take this path? What does it say about us? This is a different kind of evolution. Not the slow one based on version control systems. This is something new. Something we haven't named yet because we're too busy to notice.

\section{The Technology Explosion}

Technologies are the ultimate hydra. Cut off one head and two more grow back.

Web apps. Mobile apps. AI systems. Each one demands different thinking, different tools, different workflows. We went from writing components to assembling them from libraries. That worked. Then we got CI/CD pipelines and coding check styles. That worked too. Yeah well, of course code review practices are still firmly grounded in the stone age with absurd tools like Gerrit. Speaking of absurdities, it would have been much better if Linus would have remained in bed on that famous day he unleashed the horrors of git on mankind. Still, good times compared to today. Now we're drowning in options. Too many choices. Too much complexity.

Switching between technology stacks isn't like switching languages. It's like learning to drive on a different planet where gravity works backwards. Companies create domain-specific languages. Junior developers spend six months learning them. Then they quit. Who wants to specialize in a proprietary language when the next job needs something completely different?

Here's the nightmare scenario: technology evolves faster than expertise can accumulate. There's always something new. Always something to learn. The cost of learning scales with age. Young developers adapt fast. Senior developers have families, mortgages, actual lives outside of work. They can't spend every evening learning the hot new framework.

Trust becomes the question. Do we trust the technology? Do we trust developers to handle technologies they barely understand? Do we trust ourselves? There's no common model anymore. UML tried to be that model. It failed. Each company has its own approach. Even inside Microsoft (Microsoft!) each division does its own thing. The explosion happens because we add new technologies without getting rid of old ones. They pile up like newspapers in a hoarder's house.

People create new technologies when they don't like existing ones. The barrier to entry is low. GitHub is free, or so they say. Frameworks are open source. We try things. Most fail. Some succeed. Success spreads fast. But this isn't innovation. It's immaturity. A field that hasn't grown up yet.

Other engineering disciplines have physical constraints. You can't build a bridge that violates the laws of physics. But software? Software doesn't care about physics. JSON became infrastructure without any corporate backing, replacing the certainly more terrible XML which did have corporate backing. That couldn't happen in civil engineering or aerospace. Maybe it happens in other fields, but they have standards. Regulations. We don't.

We build products that change society. Social media. Trading algorithms. Medical devices. But we have no framework for accountability. Other engineering disciplines do. Civil engineers go to jail when bridges collapse. Software engineers? We release patches and move on.

Total luxury made us spoiled: We get multiple attempts at building software. We can prototype. We can iterate. We deal with pure abstraction. Poets building castles in the air, like Brooks said \cite{Bro95a}. How many shots does one have at building a bridge?

Even medium-sized software projects are more complex than the most elaborate physical objects humans have ever built. Think about that for a minute. Really think about it.

\section{Education and Skills}

What should we teach students?

AI-aware skills are essential now. But what do we stop teaching to make room? Programming teaches problem-solving. That's valuable. But do students need to write code themselves anymore? Or should we teach them how to understand and evaluate code that AI generates?

China is teaching AI from kindergarten. Kindergarten. Five-year-olds learning about neural networks. The shift is happening early, shaping how an entire generation thinks.

The SWEBOK guide \cite{Was24a}, the Software Engineering Body of Knowledge, keeps evolving, but slowly, so incredibly slowly, based on cumbersome work by excellent people. Like endlessly grinding a FromSoftware game. Still, new knowledge areas appear. Operations. Security. AI everywhere. Agile. DevOps. Privacy and ethics need entire courses now. Domain expertise comes later. First, you teach foundations. The engineering mindset. But how do you teach mindset? Projects help. Studying past failures would help, learning from the disasters, the projects that crashed and burned. Personal knowledge matters. Tacit knowledge that can't be written down in textbooks. SWEBOK gives you the baseline, but nobody reads the SWEBOK. Experience gives you everything else.

Students learn programming through a hundred different paths now. Different schools. Different curricula. Bootcamps. Self-taught. Online courses. There's no standard path anymore. Some people want standardized exams. Like a driving license. Like bar exams for software engineers. 

\section{Research and Community}

The software engineering research community is in crisis. Publishing is harder. Research is harder. Organizing conferences is harder. Everything is harder. Not for everyone of course, what are you on about? Just think, there’s people out there pushing 30+ papers to ICSE and FSE on a regular basis. Genius or just raw manpower? Misdemeanor or just a crime? Whatever it is, it's happening in broad daylight. Emerging role models, in any case. Consequences. Who do the young look up to?

This reminds me of the ending of the short story “Children on a Country Road” by Franz Kafka, which goes like this:“I was making for that city in the south of which it was said in our village: “There you’ll find weird folk! Just think, they never sleep!” “And why not?” “Because they never get tired.” “And why not?” “Because they’re fools.” “Don’t fools get tired?” “How could fools get tired!”

Doesn't concern me, says the person in the ivory tower. Naive. Any action, including inaction, has consequences. Sartre was right all along, indeed: "Hell is other people" \cite{Sar45a}.

Conference rankings create fuzzy boundaries. Papers migrate between venues, looking for acceptance like homeless people looking for shelter. “I’m CoreA+, oh you’re only CoreB? I’m not talking to you!” MSR (Mining Software Repositories) evolved beyond its original purpose. Went from a cool idea to becoming a bureaucrat flapping the Empirical Software Engineering standards document in your face. Accountants. Self entitled vigilantes. MSR now just another generic software engineering conference staring at the barred doors of CSRankings. Nobody likes CSRankings. Everybody uses CSRankings. Bibliometrics, the nicotine of academics and those who think that research(ers) can and should be managed. Count the beans, ignore the farts.

Ironic. We’re locked in a cell, the keys in our pockets, and complain about the prison food.

The field fragments. Software engineering is dying and reincarnating as multiple disciplines. Only a fraction of software follows modern practices. Not everything is Google or Facebook. Oh, now it’s called Meta. Anyway, VR doesn’t work, Mark, get over it.

But somewhere, someone is still maintaining COBOL. Someone is still wrestling with code that predates version control. This applies to data, too. Data is software.

We're hypothesis-driven. We build tools. We watch how people use them. Laws should emerge from wisdom and observation. But we don't observe long enough. We don't accumulate wisdom. Political regulation is coming, AI regulation, data privacy regulation, and we're not prepared.

We can provide analysis in context. Technique X works better in context Y. Not that technique X is universally best. With AI, this gets more complex. Alignment between what developers need and what tools provide matters. Research should improve developer lives, not make them harder.

We need to work closer with industry. Industry cares because money matters. When good ideas emerge, we need validation paths. But the noise is loud. Hard to identify signal. Hard to maintain overview. We have the ability to think and experiment. We should use it better.

\section{Epilogue - Next Stop: Purgatory}

Software engineering is changing. AI is part of it, but not all of it. The field fragments. Specializes. Accelerates past its ability to learn. Technologies multiply. Standards vanish into thin air. The field is still young. It serves society in a hundred different ways. It looks simple from the outside but requires massive infrastructure. Maintenance matters. Modification matters. Our community will transform. Will specialize. Must specialize to survive.

Three questions surface, like bodies in a lake: Can software makers and trained engineers coexist? Can we slow down enough to learn from our mistakes? Can we navigate the technology explosion without losing our way? No clear answers. The future is vague. What timeframe matters? One year? Five? Twenty? Different horizons give different answers, like fortune tellers reading different cards. One thing is certain: humans remain essential. We create solutions to problems. That's not going away. Yeah, we’re also good at creating problems in need of solutions. Still, we need to understand what we build. We need to learn from failures. We need proper training.

Is this still software engineering? Maybe. Maybe not. Maybe it becomes something new. Maybe it fragments into a dozen successor disciplines. Maybe that's okay. Maybe that's inevitable.

The critical component remains human. People who think and build. People who learn and teach. We're the bridge between tools and outcomes. Machines can't navigate emerging problems. We can. Until AGI descends upon us like some vengeful archangel. Ain't gonna happen, Sam.

We need wisdom. We need sustained observation. We need time to think. The field moves fast. Too fast. We need to push toward the limit of our capacity. Not sit back and coast.

Not emit garbage and call it progress.

\section{Acknowledgments}

FUSE 2024 attendees: Takumi Akazaki, Carmen Armenti, Venera Arnaoudova, Sebastian Baltes, Stefano Campanella, Oscar Chaparro, Anthony Cleve, Mattia Fazzini, Marco Gerosa, Jesus Gonzalez-Barahona, Jin Guo, Sonia Haiduc, Ahmed Hassan, Hideaki Hata, Shinpei Hayashi, Takashi Ishio, Yutaro Kashiwa, Takashi Kobayashi, Masanari Kondo, Raula Gaikovina Kula, Wing Lam, TLN, Andrian Marcus, Roberto Minelli, Kevin Moran, Nicole Novielli, Martin Robillard, Gregorio Robles, Shinobu Saito, Alexander Serebrenik, Margaret-Anne Storey, Christoph Treude, Hironori Washizaki, Thomas Zimmermann. It was a pleasure. You're welcome.

\bibliography{inferno}
\bibliographystyle{abbrvurl}
\end{document}